# Vibration of cylindrical shells: Design criteria for transition from shell modes to beam modes

M. Khademi-kouhi, M. Shakouri*

*Department of aerospace engineering, Semnan University, P.O. Box 35131-19111, Semnan, Iran.*

**Abstract**

The dynamic behavior of the cylindrical shell can be predicted by more simplified beam models for a wide range of applications. The present paper deals with finding design conditions in which the cylindrical shell performs like a beam. Employing the Hamilton's principle, the governing equations are obtained using both Donnell-Mushtari and Flugge shell theories and the analytical solution is obtained for long cylinders with simply-supported boundary conditions at both ends. Then, by equalizing the shell and beam vibration frequencies, the shell-to-beam transition conditions are obtained for both theories. To account the effects of shear distortion and rotatory inertia of the shell, the finite element method is applied to find the best transition conditions with less approximating assumptions. Finally, the effects of boundary conditions on the transition parameters as well as the frequency response are studied. The obtained conditions simply define that if the shell can be assumed as a beam in any specific geometrical and material conditions.

## 1 Introduction

Cylindrical shells are one of the most working structures in wide range of engineering applications including spacecraft, pipes and ducts and storage tanks. During operation, the cylindrical shells may subject to various loading conditions, which alter the complex problem of dynamic response (i.e. natural frequency and mode shapes) of the shell structure.

Depending on the circumferential deformations, the mode shapes of a typical cylinder can be divided in three forms, including axisymmetric (breathing), beam and self-balancing modes [1] which is shown in Fig. 1. The parameter "*n*" depicts the circumferential wave number which will be presented later. The axisymmetric mode ($n=0$) describes a constant radial change in circumferential direction, which means that the circular cross-sections remains circle with the same origin. The beam mode ($n=1$) describes the behavior of the cylindrical shell if it assumed as a circular cross-section beam. In other words, in the beam mode, the circular cross-section remains constant and only the origin of the circle fluctuates. The self-balancing modes are when the shape of circular-cross section changes as seen in Fig. 1. In many practical cases, such as low-frequency vibration of pipes in industrial installations, the beam-like modes are of interest. In addition, the dynamic behavior of the cylindrical shell can be satisfactorily predicted by more simplified beam models for a wide range of applications. Such approximations are very useful when one wishes to handle the complex problem of the response of a spacecraft to dynamic loading. [2]

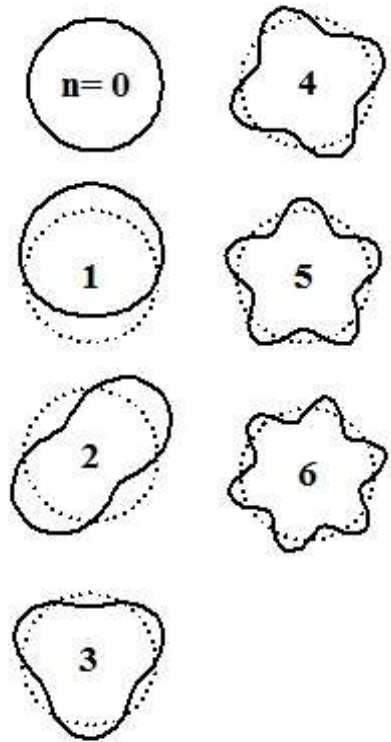

**Fig. 1.** Shell modes ($n \neq 1$) and beam mode ($n=1$) for cylindrical shells

In recent decades, considerable efforts have been devoted to the use of various shell theories for precisely prediction of the dynamic behavior of cylindrical shells [3-5]. Since the shell equations should be tolerated for conditions with several circumferential waves, the complex (usually eighth-order equations) are dealt for any shell problem. However, for long and relatively thick cylinders, the minimum natural frequency usually corresponds to the axisymmetric and beam-type modes [3], which can be satisfactorily predicted by the use of simplified beam model for a wide variety of cases.

There are some investigations attempting to analyze the complex shell equations using beam approximations. Forsberg [2] studied the breathing and beam-type vibrations of cylindrical shells applying beam approximations and compared the results with exact solutions from Flugge's shell equations. Soedel [6] used the beam functions to approximate the vibration of cylindrical shells and Farshidianfar et al. [7] employed the same approach for extracting the modal amplitudes of a long cylindrical shells. Olizadeh et al. [8, 9] studied the free vibration of simply supported cylindrical shells with four different shell theories, employing beam approximate functions and presented the effects of different parameters on mode shapes and natural frequencies. Sakar and Sonti [10] studied the vibration of an infinite fluid-filled cylindrical shell in the beam mode and found that at low frequencies, the cylindrical shell in the beam mode acts as Timoshenko beam. Pavlov and Kuptsov [11] investigated the vibrations of a rotating cylindrical shell with noncircular cross-section in beam mode. Vinson [12] used the Rayleigh-Ritz technique and Green's functions to study the beam-type vibration of cylindrical shells. Kumar et al. [13] investigated the natural frequencies of thin cylindrical shells including radial loads. Lopatin and Morozov [14] studied the axisymmetric ($n=0$) vibration of orthotropic cylindrical shell with end disks by the use of clamped-clamped beam approximating functions for deflection and axial displacement. Shen et al. [15] studied the beam-mode dynamic characteristics and stability cantilevered functionally graded fluid-conveying shells. Wang et al. [16] studied the dynamic behavior of multiwall carbon nanotubes in axisymmetric and beam modes. Fazzolari [17] proposed a Ritz formulation for the free vibration of thin-walled isotropic and composite structures with the capability of detecting the shell-like modes. Winfield et al. [18] studied the vibration of a long thick laminated conical tube using the beam approximation.

Although the simplified beam models to study the shell structures are frequently used, there are a few information about the conditions in which the beam-type modes are dominant. In other words, the transitions conditions from shell modes (i.e. axisymmetric and self-balancing modes) to beam modes needs to be more emphasized. In the most of previous papers, some specific conditions are considered and the shell equations are solved using beam approximations. For example, Blaauwendraad and Hoefakker [1] made the same procedure for static analysis of cylindrical shells and defined the conditions in which the shell acts as a beam. (see section 9 of Ref. [1]). Forsberg [2] considered a long and thin shell ($L/R>10$ and $R/h>10$) and made some discussion outlining the limitations of these approximations. To the best knowledge of the authors, there is no published condition to specify the transition condition from shell and axisymmetric modes to the beam modes.

In the present paper a design criterion is suggested in which the circular cylindrical shell behaves like a beam. For this purpose, the natural frequencies are obtained using both linear shell and beam equations. The differential equations are obtained using both Donnell-Mushtari and Flugge shell theories and the analytical solution is obtained for long cylinders with simply-supported boundary conditions at both ends. The simple shell-to-beam transition conditions are then obtained for both theories. Since the effects of shear distortion and rotatory inertia of the shell have been neglected in thin shell theories, the finite element method is used to find the best transition conditions with less approximating assumptions. Finally, the effects of boundary conditions on the transition parameters as well as the frequency response are studied. The obtained conditions simply define that if the shell can be assumed as a beam in any specific geometrical and material conditions.

## 2 Analytical solution for transition condition from shell to beam modes

### 2.1 *Displacements and strains*

A typical isotropic cylindrical shell with ($x,\theta,z$) coordinates is considered and shown in Fig. 1, where $x,\theta,z$ are longitudinal, circumferential and radial coordinates, respectively. Accordingly, the parameters $R$, $L$ and $h$ are the radius, length and thickness of the shell. The deflection of the shell's middle surface along $x$, $\theta$ and $z$ directions are denoted by $u,v$ and $w$, respectively. The Kirchhoff hypothesis cause the displacement field ($u, v, w$) to be

$$u(x,\theta,z,t) = u_0(x,\theta,t) - z\frac{\partial w_0(x,\theta,t)}{\partial x}$$

$$v(x,\theta,z,t) = v_0(x,\theta,t) - \frac{z}{R}\frac{\partial w_0(x,\theta,t)}{\partial \theta} \tag{1}$$

$$w(x,\theta,z,t) = w_0(x,\theta,t)$$

where ($u_0, v_0, w_0$) represent the mid-plane displacements along $x$ and $\theta$ directions, respectively.

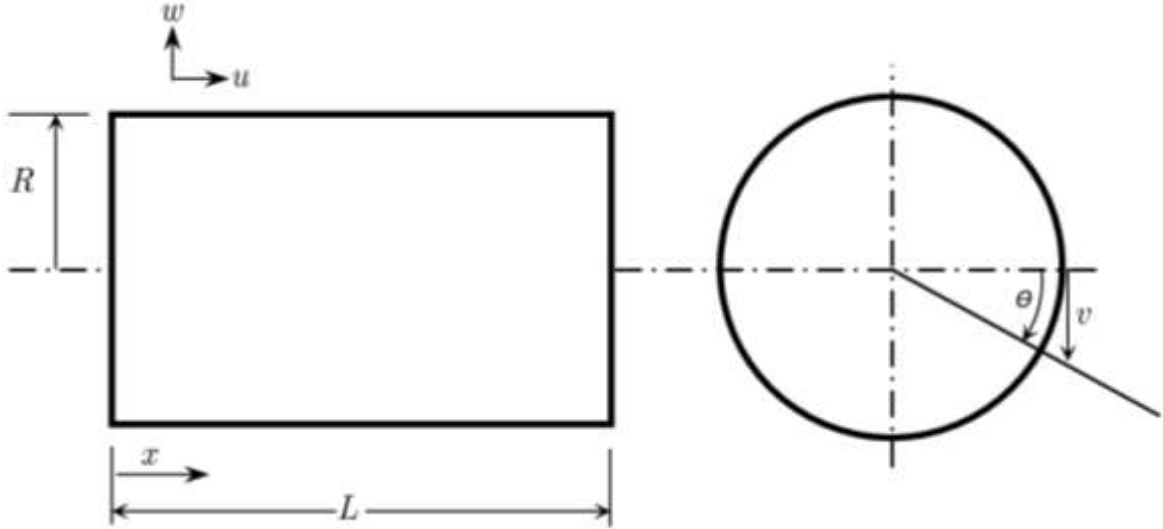

**Fig.1.** Geometry of a typical cylindrical shell.

Using the thin shell theory of Donnell-Mushtari, the strains can be written as[19]

$$\begin{Bmatrix} e_x \\ e_\theta \\ e_{x\theta} \end{Bmatrix} = \begin{Bmatrix} \varepsilon_x \\ \varepsilon_\theta \\ \gamma_{x\theta} \end{Bmatrix} + z \begin{Bmatrix} \kappa_x \\ \kappa_\theta \\ \kappa_{x\theta} \end{Bmatrix} \tag{2}$$

$$\begin{Bmatrix} \varepsilon_x \\ \varepsilon_\theta \\ \gamma_{x\theta} \end{Bmatrix} = \begin{Bmatrix} \dfrac{\partial u}{\partial x} \\ \dfrac{1}{R}(\dfrac{\partial v}{\partial \theta} + w) \\ \dfrac{1}{R}\dfrac{\partial u}{\partial \theta} + \dfrac{\partial v}{\partial x} \end{Bmatrix}, \quad \begin{Bmatrix} \kappa_x \\ \kappa_\theta \\ \kappa_{x\theta} \end{Bmatrix} = \begin{Bmatrix} -\dfrac{\partial^2 w}{\partial x^2} \\ -\dfrac{1}{R^2}\dfrac{\partial^2 w}{\partial \theta^2} \\ -\dfrac{2}{R}\dfrac{\partial^2 w}{\partial x \partial \theta} \end{Bmatrix} \tag{3}$$

The parameters ($\varepsilon_x, \varepsilon_\theta, \gamma_{x\theta}$) are membrane strains, and ($\kappa_x, \kappa_\theta, \kappa_{x\theta}$) are the curvatures.

## 2.2 Constitutive relations

For isotropic materials, the constitutive relations can be shown as

$$\begin{Bmatrix} \sigma_x \\ \sigma_\theta \\ \sigma_{x\theta} \end{Bmatrix} = \begin{bmatrix} Q_{11} & Q_{12} & 0 \\ Q_{12} & Q_{22} & 0 \\ 0 & 0 & Q_{66} \end{bmatrix} \begin{Bmatrix} e_x \\ e_\theta \\ e_{x\theta} \end{Bmatrix} \qquad (4)$$

$$Q_{11} = Q_{22} = \frac{E}{1-v^2}, \qquad Q_{12} = \frac{vE}{1-v^2}, \qquad Q_{66} = \frac{E}{2(1+v)} \qquad (5)$$

where $E$, $G$ and $v$ are Young's modulus, shear modulus and Poisson's ratio, respectively.

*2.3  Governing equations for vibration of cylindrical shells*

The governing equations of isotropic cylindrical shells are derived by the use of Hamilton's principle as

$$\int_0^T (\delta U + \delta V - \delta K)\, dt = 0 \qquad (6)$$

where $\delta U$, $\delta V$ and $\delta K$ are virtual strain energy, virtual potential energy and virtual kinetic energy, respectively and defined as

$$\begin{aligned}
\delta K &= \int_V \rho(\dot{u}_1 \delta \dot{u}_1 + \dot{v}_1 \delta \dot{v}_1 + \dot{w}_1 \delta \dot{w}_1)\, dV \\
&= \int_A \int_{-h/2}^{h/2} \rho[(\dot{u} - z\frac{\partial \dot{w}}{\partial x})(\delta \dot{u} - z\frac{\partial \delta \dot{w}}{\partial x}) \\
&\qquad + [\dot{v} + \frac{1}{R} z(\dot{v}\cos\alpha - \frac{\partial \dot{w}}{\partial \theta})][\delta \dot{v} + \frac{1}{R} z(\delta \dot{v}\cos\alpha - \frac{\partial \delta \dot{w}}{\partial \theta})] + \dot{w}\delta \dot{w}] R\, dx\, d\theta\, dz \\
&= \int_A \{I_0 (\dot{u}\delta \dot{u} + \dot{v}\delta \dot{v} + \dot{w}\delta \dot{w}) + \frac{1}{2} I_2 \delta[(\frac{\partial \dot{w}}{\partial x})^2 + \frac{1}{R^2}(\dot{v}\cos\alpha - \frac{\partial \dot{w}}{\partial \theta})^2]\} R\, dx\, d\theta
\end{aligned} \qquad (7)$$

$$\begin{aligned}
\delta U &= \int_V \sigma_{ij} \delta \varepsilon_{ij}\, dV = \int_A \int_{-h/2}^{h/2} \sigma_{ij} \delta \varepsilon_{ij} R\, ds\, d\theta\, dz \\
&= \int_A [N_x \delta \varepsilon_x + M_x \delta \kappa_x + N_\theta \delta \varepsilon_\theta + M_\theta \delta \kappa_\theta + N_{x\theta} \delta \varepsilon_{x\theta} + M_{x\theta} \delta \kappa_{x\theta}] R\, ds\, d\theta
\end{aligned} \qquad (8)$$

$$\delta V = \int_\Gamma [\hat{N}_x \delta u + \hat{T}_x \delta v + \hat{S}_s \delta w + \hat{M}_x \delta(\frac{\partial w}{\partial x})] R\, d\theta \qquad (9)$$

where $\rho$ is the density and $I_i$s are the mass inertias defined as

$$I_i = \int_{-h/2}^{h/2} \rho z^i\, dz \qquad (i = 0, 2) \qquad (10)$$

In addition, parameters $\hat{N}_x, \hat{T}_x, \hat{S}_x, \hat{M}_x$ are stress resultants due to applied axial load, and $(N, M)$ are stress resultants measured per unit length and defined as

$$\begin{bmatrix} N_x \\ N_\theta \\ N_{x\theta} \\ M_x \\ M_\theta \\ M_{x\theta} \end{bmatrix} = \int_{-h/2}^{h/2} \begin{bmatrix} \sigma_x \\ \sigma_\theta \\ \sigma_{x\theta} \\ z\sigma_x \\ z\sigma_\theta \\ z\sigma_{x\theta} \end{bmatrix} dz \tag{11}$$

Substitting Eqs.(3), (4) and (11) into Eqs.(7)-(9), neglecting $I_2$ due to thin shell assumptions and then imposing all into Eq. (6), we have [3]

$$\begin{aligned}
\delta u &: \frac{\partial N_x}{\partial x} + \frac{1}{R}\frac{\partial N_{x\theta}}{\partial \theta} = I_0 \frac{\partial^2 u}{\partial t^2} \\
\delta v &: \frac{1}{R}\frac{\partial N_\theta}{\partial \theta} + \frac{\partial N_{x\theta}}{\partial x} + \frac{1}{R}\frac{\partial M_{x\theta}}{\partial x} + \frac{1}{R^2}\frac{\partial M_\theta}{\partial \theta} = I_0 \frac{\partial^2 v}{\partial t^2} \\
\delta w &: -\frac{1}{R}N_\theta + \frac{\partial^2 M_x}{\partial x^2} + \frac{1}{R}\frac{\partial^2 M_{x\theta}}{\partial x \partial \theta} + \frac{1}{R}\frac{\partial^2 M_{x\theta}}{\partial x \partial \theta} + \frac{1}{R^2}\frac{\partial^2 M_\theta}{\partial \theta^2} = I_0 \frac{\partial^2 w}{\partial t^2}
\end{aligned} \tag{12}$$

*2.4 Vibration analysis of long cylindrical shell*

In the case that the longitudinal wavelength is long, the solution functions can be considered as [3]

$$\begin{aligned}
u(x,\theta,t) &= U \cos n\theta\, e^{i\omega t} \\
v(x,\theta,t) &= V \sin n\theta\, e^{i\omega t} \\
w(x,\theta,t) &= W \cos n\theta\, e^{i\omega t}
\end{aligned} \tag{13}$$

where $(U,V,W)$ are the amplitudes, $n$ is the circumferential wave number and $\omega$ is the frequency of the cylindrical shell. Applying Eq.(13) in governing equations yields a set of algebraic equations as [3]

$$\begin{bmatrix} \frac{1-v}{2}n^2 - \frac{\rho(1-v^2)R^2\omega_s^2}{E} & 0 & 0 \\ 0 & n^2 - \frac{\rho(1-v^2)R^2\omega_s^2}{E} & n \\ 0 & n & (1+kn^4) - \frac{\rho(1-v^2)R^2\omega_s^2}{E} \end{bmatrix} \begin{Bmatrix} U \\ V \\ W \end{Bmatrix} = \begin{Bmatrix} 0 \\ 0 \\ 0 \end{Bmatrix} \tag{14}$$

The subscript "s" denotes the frequency formula obtained from shell equations and

$$k = \frac{h_e^4}{12} \tag{15}$$

in which $h_e$ is the equivalent thickness, defined as

$$h_e = \sqrt{\frac{h}{R}} \tag{16}$$

It can be seen from Eq.(15) that the motion in axial (longitudinal) direction is completely uncoupled from the other two modes. Therefore, finding the roots of the second order determinant for non-axial modes, arising from Eq.(15) yields the frequencies of the long cylindrical shells with Donnell-Mushtari shell theory as

$$\omega_s^2 = \frac{E}{2\rho(1-v^2)R^2}[(1+n^2+kn^4) \mp \sqrt{(1+n^2+kn^4)^2 - 4kn^6}] \quad n \neq 0 \tag{17}$$

Following the same procedure, the frequency parameter for circumferential modes of long cylindrical shell according to Flugge shell theory can be extracted as [3]

$$\omega_s^2 = \frac{E}{2\rho(1-v^2)R^2}[(1+n^2+kn^4) \mp \sqrt{(1+n^2)^2 - 2kn^6}] \quad n \neq 0 \tag{18}$$

### 2.5 Vibration of the simply-supported beam

The vibration characteristics of an Euler-Bernoulli beam with simply-supported boundary conditions at both ends can be easily found in any reference book (see for example Craig and Kurdila[20]) as

$$\omega_b = (\frac{m\pi}{L})^2 \sqrt{\frac{EI}{\rho A}} \tag{19}$$

where m is the longitudinal half wave number and the subscript "b" denotes the frequency formula obtained from beam equation, $I$ is the second moment of area and $A$ is the cross-section area. For a thin tube, the Eq.(20) can be represented as

$$\omega_b = R(\frac{m\pi}{L})^2 \sqrt{\frac{E}{2\rho}} \tag{20}$$

### 2.6 Transition condition from shell modes to beam modes

The transition conditions in which, the vibration of the cylindrical shell converts from shell-modes to the beam modes can be easily extracted when the frequencies from shell equations (i.e. either Eq. (18) or Eq.(19)) be equalized with the frequencies obtained from beam solution (Eq.(21)). Considering the beam mode ($n=m=1$), neglecting higher terms of $k$ and performing some mathematical simplification, the transition conditions can be stated as

$$L_e h_e = C \sqrt[4]{(1-v^2)} \tag{21}$$

where $L_e$ id the equivalent length defined ass

$$L_e = \frac{L}{R} \tag{22}$$

and *C* is the constant depends on the applied shell theory. For Donnel-Mushtari and Flugge shell theories the critical value of the C constant can be extracted as

$$C_{cr} = 5.847 \quad \text{Donnell-Mushtari theory}$$
$$C_{cr} = 5.284 \quad \text{Flugge theory} \tag{23}$$

The above relation depicts that the shell acts as a beam when the geometrical and material conditions of the shells are in the way that $C>C_{cr}$, the shell behaves like a beam and vice versa.

## 3 Shell-to-beam mode transition using finite element analysis

The aim of this section is to find the transition conditions from shell-like modes to beam-like modes using finite element analysis. To this end, the finite element procedure is described first and then, the effects of material properties on the transition condition are studied. Finally, the value of the $C_{cr}$ is obtained and the results are discussed in detail.

### *3.1 Finite element analysis*

Since the shell theories are simplified with some assumptions, the finite element (FE) analysis is employed to find the best conditions (with less error) for transition from shell to beam modes. A cylindrical shell made from Aluminum (*E*=70 Gpa and *v*=0.3) with different radii and thicknesses is considered. By changing the length, thickness and radius of the shell, the minimum value of the geometrical properties with beam-like mode (i.e. *n*=1) can be obtained. In the next step, the mechanical properties and boundary conditions are changed and the effects of these parameters on the shell-to-beam transition are studied.

The Finite element analysis is conducted by ABAQUS software with four nodes shell element S4R that has six degrees of freedom at each node, three translational displacements in the nodal directions and three rotational displacements about the nodal axes. S4R is a linear element formulation, reduced-integration, and hourglass control. The eigenvalue problem is solved using LANCZOS method.

Generally, the result of the FE analysis is affected by number of elements. Hence, the convergence of the analysis results versus the number of total elements are studied. In order to select appropriate mesh size, a compromise between time and number of elements is needed. In other words, by increasing the number of elements, there is no specific change in the solution and indeed the cost of computation can overcome the changes of the solution. The final FE model of the structure is composed of 7560 elements and 7860 nodes.

### *3.2 Effects of elastic modulus*

In this section, the effect of different values of Young moduli and Poisson ratios on the transition condition is investigated. It should be noted that from previous studied, we know that the elastic modulus has no effect on the transition conditions. However, to ensure from the correctness and accuracy of the results, the Young modulus is changed and the its effects on the results are studied.

To study the elastic modulus, a cylindrical shell with constant radius $R$=10mm and thickness $h$=1mm with $\rho$=2700Kg/m$^3$ and $v$=0.3 is considered. The values of elastic modulus are varied from 3 to 300 GPa and the minimum value of the shell length, in which, the beam-like mode ($n$=1) occurs is obtained. The results are shown in Table 1. It can be seen that the elastic modulus has no effect on transition condition.

**Table 1.** Effects of Young's modulus on equivalent length and thickness.

| $E$ (GPa) | 3 | 70 | 100 | 150 | 180 | 210 | 240 | 300 |
|---|---|---|---|---|---|---|---|---|
| $h_e$ | 0.3162 | 0.3162 | 0.3162 | 0.3162 | 0.3162 | 0.3162 | 0.3162 | 0.3162 |
| $L_e$ | 14 | 14 | 14 | 14 | 14 | 14 | 14 | 14 |

### 3.3 Effects of Poisson's ratio

From previous studies, it can be concluded that the shell-to-beam mode transition condition is proportional with the forth root of the (1-$v^2$), as seen in Eq.(21). However, to ensure from the Poisson's ratio effects, the above-mentioned aluminum cylinder (i.e. $R$=10mm, $h$=1mm, $\rho$=2700Kg/m$^3$ and $E$=70GPa) is considered. The values of Poisson's ratio are varied from 0.18 to 0.4 and the minimum length in which the beam mode occurs is obtained.

**Table (2)** Effects of Poisson's ratio on equivalent length and thickness of transition condition from shell to beam modes.

| $v$ | 0.18 | 0.22 | 0.26 | 0.3 | 0.35 | 0.4 |
|---|---|---|---|---|---|---|
| $h_e$ | 0.3162 | 0.3162 | 0.3162 | 0.3162 | 0.3162 | 0.3162 |
| $L_e$ | 14 | 14 | 14 | 13 | 13 | 13 |

Table 2 shows the values of equivalent length and equivalent thickness for different values of Poisson's ratio. It can be concluded that

$$L_e h_e \propto \sqrt[4]{(1-v^2)} \qquad (24)$$

## 4 Shell-to-beam transition condition

The FE analysis was performed for a wide range of cylindrical shells with 0.001<$R$<1 m and 0.001<$h/R$<0.1. In each specific $R$ and $h$, the values of the shell length are changed and the minimum value at which the beam-like mode is occurred is obtained. The results of FE analysis for all the geometries, along with Donnel-Mushtari and Flugge results (Eq. 22) are plotted in the $L_e$-$h_e$ space, as shown in Fig. 3. It can be seen that, due to the simplifications in Donnel-Mushtari and Flugge theories, the results of these theories are a bit different with FE results. In other words, the shell-to-beam transition condition is a more than the present results.

To find the best values for transition condition, the best curve fitted to the FE results is obtained. The curve like the Eq. (23) is applied and the $C_{cr}$ is obtained as

$$C_{cr} = 4.49 \tag{25}$$

Equation (25) denotes the transition condition from shell-like modes to beam-like modes obtained from FE analysis. In other words, if the geometrical and material properties of the shell are in the way that the value of $L_e h_e / \sqrt[4]{(1-\nu^2)}$ is more than $C_{cr}$, the shell can be assumed as a beam.

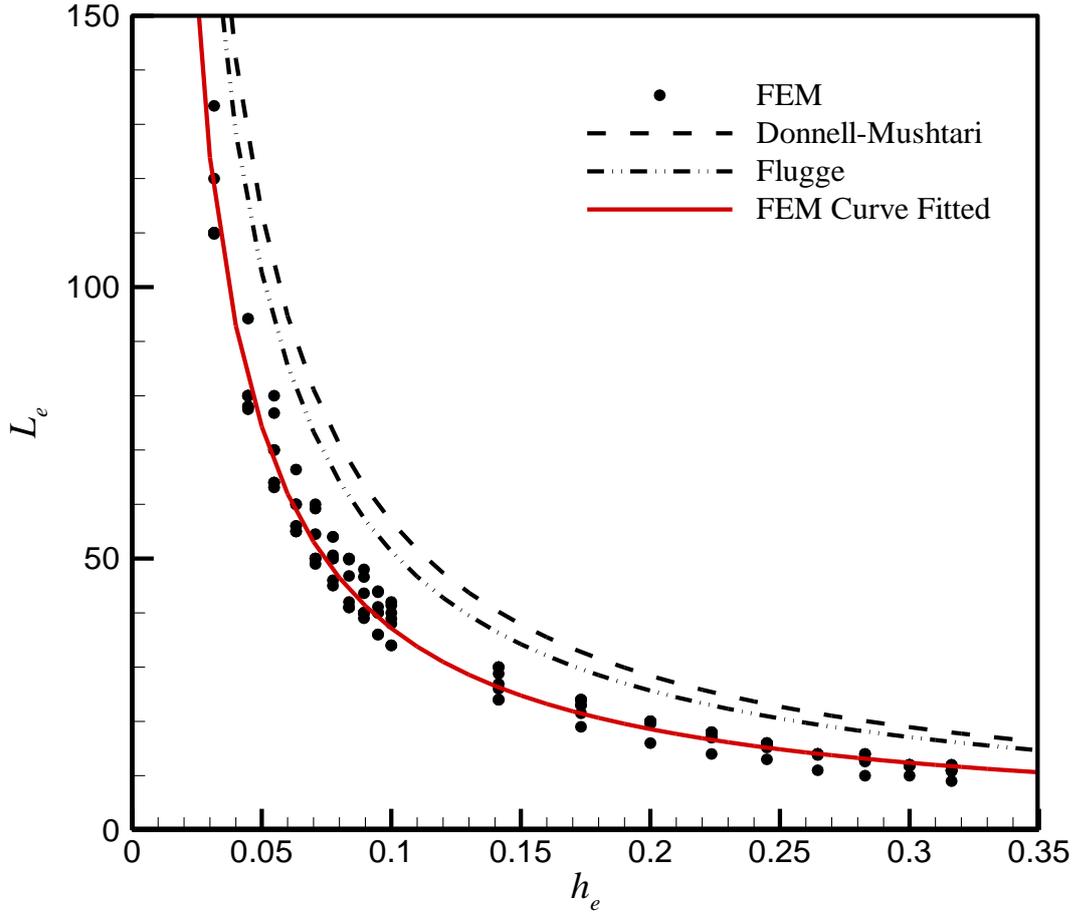

**Fig. 3:** transition condition from shell to beam modes, obtained from Donnel-Mushtari, Flugge and FEM.

## 5   Effects of boundary conditions

In this section, the effects of different boundary conditions on the transition conditions of cylindrical shell is studied. Three boundary conditions, including free-free (F-F), simply supported-simply supported (SS), and clamped-clamped (CC) are considered and the results are shown in Fig. 4. As can be seen, the results of different boundary conditions are very close to each other. As a result, the effects of boundary conditions on the transition condition from shell-like modes to beam-like modes are negligible. It was not unexpected, because the beam mode usually occurs for relatively long cylinders and the effects of boundary conditions on the dynamic behavior of the shells decrease with increase in the shell length.

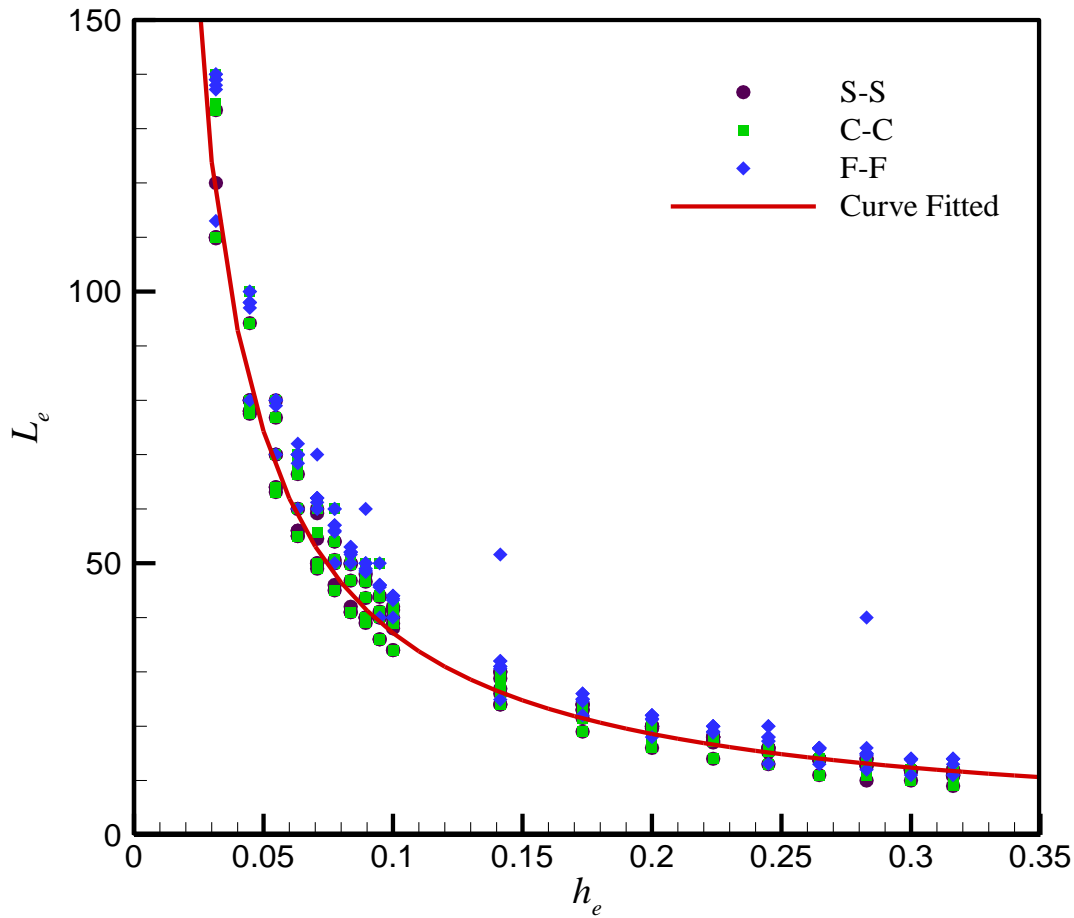

**Fig. 4.** Effects of boundary conditions on shell-to-beam transition condition.

## 6  Conclusion

In this paper a transition condition from shell-like modes (i.e. axisymmetric and self-balancing modes) to beam-like modes is obtained. If the geometrical and material properties of the shell meet this condition, the cylindrical shell behaves like a beam and the governing equations of beam can be applied. This is a valuable achievement, since the dynamic behavior of the beam can be easily studied by simpler governing equations. The Donnell-Mushtari and Flugge shell theories are employed to derive the governing equations and the transition condition is achieved by equalizing the natural frequencies obtained from shell and beam theories. The simple shell-to-beam transition conditions are then obtained and to account the effects of shear distortion and rotatory inertias, the more accurate relation is extracted using finite element analysis. In addition, the effects of boundary conditions on the transition parameters are studied. The obtained transition relation simply defines the condition in which the shell can be assumed to behave like a beam.